\documentclass[12pt]{iopart}%\documentclass[12pt]{iopart}
\usepackage{graphicx}%\usepackage{iopart}
\usepackage{amssymb}
\usepackage{iopams}
\usepackage{amsthm}
\usepackage{latexsym}
\newcommand{\bdm}{\begin{displaymath}}
\newcommand{\edm}{\end{displaymath}}
\newcommand{\be}{\begin{equation}}
\newcommand{\ee}{\end{equation}}
\newcommand{\bea}{\begin{eqnarray}}
\newcommand{\nn}{\nonumber}
\newcommand{\eea}{\end{eqnarray}}
\newcommand{\bn}{\boldsymbol{\nabla}}

\begin{document}
\title[On the Multipole Moments of Axisymmetric Electrovacuum Spacetimes]
{Corrections and Comments on the Multipole Moments of Axisymmetric Electrovacuum Spacetimes}

\author{Thomas P. Sotiriou\footnote[1]{e-mail:tsotiri@phys.uoa.gr} and Theocharis A. Apostolatos
\footnote[2]{e-mail:tapostol@cc.uoa.gr}
}

\address{ Section of Astrophysics, Astronomy, and Mechanics\\
Department of Physics \\
National and Kapodistrian University of Athens\\
Panepistimiopolis, Zografos GR-15783, Athens, Greece}

\begin{abstract}
Following the method of Hoenselaers and Perj\'{e}s we present a new corrected and dimensionally consistent
set of multipole gravitational and electromagnetic moments for stationary axisymmetric spacetimes.
Furthermore, we use our results to compute the multipole moments, both gravitational and electromagnetic,
of a Kerr-Newman black hole.
\end{abstract}

%\pacs{}

%%%%%%%%%%%%%%%%%%%%%%
\section{Introduction}
\label{sec:1}%%%%%%%%%
After Ernst's first paper on reformulating Einstein's equations for stationary and axisymmetric spacetimes
(i) without \cite{Ernst1}, and (ii) with electromagnetic field \cite{Ernst2}, Geroch and Hansen \cite{Geroch, Hansen}
 defined the multipole moments of  static and stationary spacetimes respectively,
by measuring the deviation of the geometry from flatness in the neighborhood of a point at infinity.
Their definition of the moments is equivalent to the one introduced by Thorne in \cite{Thorne}.
Later, Simon and Beig \cite{Simon1} reconstructed the moments from the series expansion of
field variables, and Simon \cite{Simon2}
generalized them so as to include electrovacuum spacetimes.
Especially the axisymmetric spacetimes that Ernst studied
are  interesting physically, and can be fully described by a
number of scalar moments, in the same way that newtonian gravitational potentials
are characterized by their mass multipole moments.

In 1990 Hoenselaers and Perj\'{e}s \cite{HoensPerj} generalized the work of Fodor et al \cite{Fodoretal}
on vacuum gravitational moments to cases where spacetime is endowed with an electromagnetic
field that retains the axisymmetry of the metric.
Although the generalization method they used was right, an incorrect intermediate
formula led to the contamination of the final expressions for
the moments with errors. In our paper we have tried to (i) determine the
root of these errors, (ii) give a few extra intermediate relations
that are used in the procedure of computing the moments, and (iii) correct the final
expressions that relate the gravitational and electromagnetic multipole moments of  spacetime with
the power series coefficients of the gravitational and electromagnetic potentials.
The final expressions for the gravitational
and electromagnetic moments may be important for people
that are looking for approximate expressions for the metric of stationary and axisymmetric sources
that depend on a few physical parameters of the source. We actually needed these expressions
when we tried to extend \cite{SotApo1}
Ryan's study \cite{Ryan} on ways to determine the moments of a compact object from
gravitational wave data analysis, to axisymmetric spacetimes with electromagnetic
fields. The dimensional inconsistency of the moments of \cite{HoensPerj},
that we were to use, drew our attention and we set out to thoroughly check the whole procedure that
Hoenselaers and Perj\'{e}s followed in their paper.

The rest of the paper is organized as follows. In Section \ref{sec:2} we briefly repeat
the basic steps of the method used, giving a few extra intermediate expressions that may be useful to
the reader. Although most intermediate steps can be found in a number of previous articles,
the plethora of different conventions and notations used by different authors, urged us to systematically present
the ones we used to calculate the moments.
In Section \ref{sec:3} we present the first five gravitational and electromagnetic moments
as functions of the power series coefficients of the gravitational and electromagnetic potentials.
A number of comments on the
terms that show up at each moment have been added at the end. In Section \ref{sec:4}
we compute once again the first five power series coefficients of the mass and  electromagnetic potentials
of a pole-dipole source as in \cite{HoensPerj}. Finally, in Section
\ref{sec:5} we use the whole technique to compute the moments of the Kerr-Newman metric,
from the knowledge of the metric along its axis of symmetry. The gravitational moments are exactly
like in Kerr metric, while the electromagnetic moments are simply $e/m$ times
the corresponding gravitational moments.

%%%%%%%%%%%%%%%%%%%%%%%%%%%%%%%%%%%%%
\section{Ernst functions and moments}
\label{sec:2}%%%%%%%%%%%%%%%%%%%%%%%%

The metric of a stationary axisymmetric spacetime  could be described by  Papapetrou's
line element \cite{Papa}
\be
\label{metric}
ds^2= -F(dt-\omega d\phi)^2+F^{-1}\left[e^{2\gamma}(d\rho^2+dz^2)+\rho^2 d\phi^2 \right],
\ee
where $F,\omega$, and $\gamma$ are functions of $\rho$ and $z$. Assuming that spacetime
is asymptotically flat, $\gamma \rightarrow 0$, $F\rightarrow 1$, and $\omega \rightarrow 0$
at infinity. As Ernst has shown,
any axially symmetric solution of the coupled Einstein - Maxwell equations
can be described by two complex functions ${\cal E}$ and $\Phi$ which are related with
the metric functions $F$ and $\omega$ and the  components of the electromagnetic potential $A_t$, $A_\phi$
(cf., Eqs.~(8,9,11,14) of \cite{Ernst2}).
The third metric function $\gamma$ can be deduced by using the two remaining
Einstein equations.

Now, instead of ${\cal E}$ and $\Phi$ we use the complex Ernst potentials $\xi,q$, which are the analogues of
newtonian gravitational potential and Coulomb potential respectively,
\bea
\label{Efunction}
{\cal E}=\frac{1-\xi}{1+\xi} ,\qquad \Phi=\frac{q}{1+\xi}.
\eea
Note that the former relation is the opposite of Ernst (cf., \cite{Ernst1,Ernst2}), and thus
our $\xi$ is the inverse of Ernst's $\xi_E$.
It is easy to show that the Einstein equation for $\cal E$ and $\Phi$ \cite{HoensPerj}
take the following form for $\xi$ and $q$
\bea
\label{Einstxi}
(\xi \xi^\ast - q q^\ast -1) \nabla^2 \xi &=&
2 (\xi^\ast \bn \xi - q^\ast \bn q )
\cdot \bn \xi \\
\label{Einstq}
(\xi \xi^\ast - q q^\ast -1) \nabla^2 q &=&
2 (\xi^\ast \bn \xi - q^\ast \bn q )
\cdot \bn q .
\eea
By $^\ast$ we denote the complex conjugate.
The Ricci tensor of the three-metric is given by \cite{HoensPerj}
\be
\label{ricci}
(\xi \xi^\ast - q q^\ast -1)^{2} R_{ij}
=
2 \Re \left(
\nabla_i \xi \nabla_j \xi^\ast -
\nabla_i q   \nabla_j q^\ast +
s_i s^\ast_j \right),
\ee
where
\be
s_i=
\xi \nabla_i q - q \nabla_i \xi.
\ee
$\nabla_i$ denotes the components of the gradient, and $\Re()$ denotes the real part.

In order to measure the moments of an asymptotically flat spacetime,
according to Geroch-Hansen procedure \cite{Geroch,Hansen},
we map the initial three-metric to a conformal one
\be
h_{ij} \rightarrow \tilde{h}_{ij}=  \Omega^2 h_{ij}.
\ee
The conformal factor $\Omega$ should satisfy the following conditions:
$ \Omega |_\Lambda=\tilde{D}_i \Omega |_\Lambda=0$ and
$ \tilde{D}_i \tilde{D}_j \Omega |_\Lambda=2  h_{ij}  |_\Lambda$, where
$\Lambda$ is the point added to the initial manifold that represents infinity.
$\Omega$ transforms $\xi$ and $q$ potentials to
\be
\tilde{\xi}=\Omega^{-1/2} \xi ,\qquad \tilde{q}=\Omega^{-1/2} q.
\ee
By a coordinate transformation
\bea
\label{transcooel}
\bar {\rho }&=&\frac{\rho }{\rho ^2 + z^2},\nn\\
\bar {z}&=&\frac{z}{\rho ^2 + z^2},\nn\\
\bar {\phi }&=&\phi
\eea
we bring infinity at the origin of the axes $(\bar{\rho},\bar{z})=(0,0)$. Then, by choosing
the conformal factor to be
\be
\Omega=\bar{r}^2=\bar{\rho}^2+\bar{z}^2,
\ee
the conformal metric in the new coordinates takes the following form
\begin{equation}
\label{metrichel}
\tilde{h}_{ij} = \left( {{\begin{array}{*{20}c}
 {e^{2\gamma }} \hfill & 0 \hfill & 0 \hfill \\
 0 \hfill & {e^{2\gamma }} \hfill & 0 \hfill \\
 0 \hfill & 0 \hfill & {\bar{\rho} ^2} \hfill \\
\end{array} }} \right),
\end{equation}
which is flat at  $\bar{r}=0$, since $\gamma|_{\bar r=0}=0$. The Ricci tensor of this conformally transformed
three-geometry in barred coordinates is
\bea
\label{riccitild}
(\bar{r}^{2} \tilde{\xi}^{\ast} \tilde{\xi}-\bar{r}^{2} \tilde{q}^{\ast} \tilde{q}-1)^{2}
\tilde{R}_{ij}&=&
2 \Re (D_{i}\tilde{\xi}D_{j} \tilde{\xi}^{\ast}-
       D_{i}\tilde{q}  D_{j}   \tilde{q}^{\ast}+
       \tilde{s}_{i}\tilde{s}_{j}^{\ast} ),
\eea
as one could verify by a tedious but straightforward calculation, with
\bea
\label{afterriccitild}
\tilde{D}_{1}&=&\bar{z}   \frac{\partial}{\partial \bar{\rho}}-
        \bar{\rho}\frac{\partial}{\partial \bar{z}}     \nn\\
\tilde{D}_{2}&=&\bar{\rho}\frac{\partial}{\partial \bar{\rho}}+
        \bar{z}   \frac{\partial}{\partial \bar{z}}+1   \nn\\
\tilde{s}_{i}&=&\bar{r}( \tilde{\xi}D_{i}\tilde{q}- \tilde{q}D_{i}\tilde{\xi}).
\eea
The $\tilde{D}_i$'s are the $\tilde{\nabla}_i$'s of \cite{HoensPerj} with their indices interchanged.
Notice that this has exactly the same form as Eq.~(\ref{ricci})
if we replace the free (not differentiated) $\xi$'s and $q$'s with $\bar{r} \tilde{\xi}$
and $\bar{r} \tilde{q}$, respectively, while the differentiated $\xi$'s and $q$'s
simply with $\tilde{\xi}$ and $\tilde{q}$, respectively.
In Eq.~(6) of \cite{HoensPerj} which was used to evaluate the  $s_i$'s, the free
$\xi$ and $q$ were replaced by $\tilde{\xi}$
and $\tilde{q}$, instead of  $\bar{r} \tilde{\xi}$
and $\bar{r} \tilde{q}$, to get the $\tilde{s}_i$'s.
This is exactly what caused the
dimensional inconsistency in the final expressions for the moments in \cite{HoensPerj}.
The field equations (\ref{Einstxi},\ref{Einstq}) keep exactly their form
in the conformal metric if expressed in the barred coordinates, by using the same replacements.

The tensorial multipole moments of the electrovacuum spacetime are computed by the recursive relation
(see \cite{Fodoretal})
\bea
\label{elmmxi}
P^{(0)}                             &=& \tilde{\xi},\nn\\
P^{(1)}_{i}                         &=& \tilde{\xi}_{,i},\nn\\
P^{(n+1)}_{i_{1}i_{2}\ldots i_{n+1}}&=& {\cal C}[\tilde{\nabla}_{i_{n+1}}P^{(n)}_{i_{1}\ldots i_{n}}-\frac{1}{2}
n(2n-1)R_{i_{1}i_{2}}P^{(n-1)}_{i_{3}\ldots i_{n+1}}],
\eea
for the geometry, and
\bea
\label{elmmq}
Q^{(0)}&=&\tilde{q},\nn\\
Q^{(1)}_{i}&=&\tilde{q}_{,i},\nn\\
Q^{(n+1)}_{i_{1}i_{2}\ldots i_{n+1}}&=&{\cal C}[\tilde{\nabla}_{i_{n+1}}Q^{(n)}_{i_{1}\ldots i_{n}}-\frac{1}{2}
n(2n-1)R_{i_{1}i_{2}}Q^{(n-1)}_{i_{3}\ldots i_{n+1}}].
\eea
for the electromagnetic field. The symbol $\tilde{\nabla}$ is used to denote the
covariant derivative in the conformal space and it should not be confused
with the same symbol used in \cite{HoensPerj}.  The operator $\cal C$ denotes the operation
``symmetrize over all free indices and take the trace-free
part''. Note that all tensors should be
evaluated at $\Lambda$ (at infinity).
Due to axisymmetry, the components of these tensorial moments are multiples of
the corresponding scalar moments which are the projection of the tensorial moments on the axis of symmetry.
Following Fodor \textit{et al} \cite{Fodoretal}, the scalar moments are defined as
\bea
\label{scalarpq}
P_n &=& \frac{1}{n!}\left. {\tilde {P}_{2...2}^{(n)} }\right|_\Lambda\nn\\
Q_n &=& \frac{1}{n!}\left. {\tilde {Q}_{2...2}^{(n)} }\right|_\Lambda.
\eea
Note that these moments have the opposite sign from the ones defined by Hansen \cite{Hansen}.
The gravitational scalar moments are then simply given by
%%%%%%%%%%
\begin{equation}
\label{pnfinal}
P_n = \frac{1}{(2n - 1)!!}S_0^{(n)},
\end{equation}
%%%%%%%%%%%%
where $S_0^{(n)}$ are computed by the recursive relations (23) of \cite{Fodoretal}.
The electromagnetic scalar moments $Q_n$ are computed from exactly the same
recursive formulae, if we replace the initial term $S_0^{(0)}$
with $\tilde {q}$ instead of $\tilde{\xi}$.
In order to use these recursive formulae one needs also the derivatives of $\gamma$
which could be expressed as linear combinations of  the components of the Ricci tensor which
is given in Eq.~(\ref{riccitild})
\bea
\label{gammaxiq}
\gamma_1 \equiv
\gamma_{,\bar{\rho}}&=&\frac{1}{2}\bar{\rho}(\tilde{R}_{\bar{\rho}\bar{\rho}}-
\tilde{R}_{\bar{z}\bar{z}}),\nn\\
\gamma_2 \equiv
\gamma_{,\bar{z}}&=&\bar{\rho}\tilde{R}_{\bar{\rho} \bar{z}}.
\eea

We end up this section by
writing both $\xi$ and $q$ as power series of $\bar{\rho}$,$\bar{z}$:
\bea
\label{expxiq}
\tilde {\xi } = \sum\limits_{i,j = 0}^\infty {a_{ij}\bar {\rho }^i\bar {z}^j},\nn\\
\tilde {q} = \sum\limits_{i,j = 0}^\infty {b_{ij}\bar {\rho }^i\bar {z}^j},
\eea
%%%%%%%%%%%%
where $a_{ij}$ and $b_{ij}$ vanish when $i$ is odd (this reflects the analyticity of both potentials
at the axis of symmetry).
Because of Einstein's  field equations
(see Eqs.~(\ref{Einstxi},\ref{Einstq}) and the  discussion for the tilded version of them
which follows after
Eq.~(\ref{afterriccitild})), the coefficients in the above power series are algebraically interrelated:
\bea
\label{abrecur1}
 \left(r+2\right)^2a_{r+2,s}&=&-(s+2)(s+1)a_{r,s+2}+{} \nn\\
 && {}+\sum_{k,l,m,n,p,g}
(a_{kl} a_{mn}^\ast-b_{kl}b^\ast_{mn})\times {} \nn\\
&& {}\times[a_{pg} (p^2 + g^2 - 4p - 5g - 2pk - 2gl -
2) + {} \nn\\
 && {}+ a_{p + 2,g - 2} (p + 2)(p + 2 - 2k) + {}\nn\\
&& {}+a_{p - 2,g + 2} (g + 2)(g + 1 - 2l)],
\eea
and
\bea
\label{abrecur2}
 \left(r+2\right)^2b_{r+2,s}&=&-(s+2)(s+1)b_{r,s+2}+{} \nn\\
 && {}+\sum_{k,l,m,n,p,g}
(a_{kl} a_{mn}^\ast-b_{kl}b^\ast_{mn})\times {} \nn\\
&& {}\times[b_{pg} (p^2 + g^2 - 4p - 5g - 2pk - 2gl -
2) + {} \nn\\
 && {}+ b_{p + 2,g - 2} (p + 2)(p + 2 - 2k) + {}\nn\\
&& {}+b_{p - 2,g + 2} (g + 2)(g + 1 - 2l)],
\eea
where $m=r-k-p$ , $0\leq k \leq r$, $0 \leq p \leq r-k$ , with $k$ and $p$ even,
and $n=s-l-g$, $0 \leq l \leq s+1$ , and $-1 \leq g \leq s-l$.
These recursive relations could built the whole power series of $\tilde{\xi}$
and $\tilde{q}$ from their value on the axis of symmetry
\bea
\tilde{\xi}(\bar\rho=0)=\sum_{i=0}^{\infty} m_i {\bar z}^i ,\qquad
\tilde{q}(\bar\rho=0)=\sum_{i=0}^{\infty} q_i {\bar z}^i.
\eea
Thus from $\tilde{\xi}(\bar{\rho}=0)$
and $\tilde{q}(\bar{\rho}=0)$, we can
read the multipole moments of spacetime.
In the next section we will write the expressions that connect the multipole moments of
both types with the $m_i$'s and $q_i$'s.

%%%%%%%%%%%%%%%%%%%%%%%%%%%%%%%%%%%%%%%%%%%%%%%%%%%%%%%%%%%%%%%%%%%%%%%%%%%%%%%%%%%
\section{The multipole moments of stationary axisymmetric electrovacuum spacetimes}
\label{sec:3}%%%%%%%%%%%%%%%%%%%%%%%%%%%%%%%%%%%%%%%%%%%%%%%%%%%%%%%%%%%%%%%%%%%%%%

Following the procedure that is outlined in Section \ref{sec:2} we compute the first five
gravitational and electromagnetic multipole moments in terms of $a_{ij}$ and $b_{ij}$.
Then the algebraic relations between  $a_{ij}$ and $b_{ij}$ are used in order to express
the moments in terms of $m_i \equiv a_{0i}$ and $q_i \equiv b_{0i}$. If we define the following
useful quantities
%%%%%%%%%%%
\bea
M_{ij}&=&m_{i}m_{j}-m_{i-1}m_{j+1},\nn\\
Q_{ij}&=&q_{i}q_{j}-q_{i-1}q_{j+1},\nn\\
S_{ij}&=&m_{i}q_{j}-m_{i-1}q_{j+1},\nn\\
H_{ij}&=&q_{i}m_{j}-q_{i-1}m_{j+1}
\eea
the gravitational moments as functions of the power series coefficients of $\tilde\xi$ along the
symmetry axis are given by
%%%%%%%%
\bea
\label{mmelfin}
P_{0}&=&m_{0},\nn\\
P_{1}&=&m_{1},\nn\\
P_{2}&=&m_{2},\nn\\
P_{3}&=&m_{3},\nn\\
P_{4}&=&m_{4}-\frac{1}{7}m_{0}^{\ast}M_{20}+\frac{1}{7}q_{0}^{\ast}S_{20}-\frac{3}{70}q_{1}^{\ast}S_{10},\nn\\
P_{5}&=&m_{5}-\frac{1}{3}m_{0}^{\ast}M_{30}-\frac{1}{21}m_{1}^{\ast}M_{20}+\frac{1}{3}q_{0}^{\ast}S_{30}+\frac{4}{21}q_{0}^{\ast}S_{21}-{}\nn\\
& &{}-\frac{1}{21}q_{1}^{\ast}S_{11}-\frac{1}{21}q_{2}^{\ast}S_{10},
\eea
%%%%%%%%%
%%%%%%%%
while the electromagnetic moments as functions of the power series coefficients of $\tilde q$ along the
symmetry axis are given by
%%%%%%%%%%%
%%%%%%%%%%%
\bea
\label{qqelfin}
Q_{0}&=&q_{0},\nn\\
Q_{1}&=&q_{1},\nn\\
Q_{2}&=&q_{2},\nn\\
Q_{3}&=&q_{3},\nn\\
Q_{4}&=&q_{4}+\frac{1}{7}q_{0}^{\ast}Q_{20}-\frac{1}{7}m_{0}^{\ast}H_{20}+\frac{3}{70}m_{1}^{\ast}H_{10},\nn\\
Q_{5}&=&q_{5}+\frac{1}{3}q_{0}^{\ast}Q_{30}+\frac{1}{21}q_{1}^{\ast}Q_{20}-\frac{1}{3}m_{0}^{\ast}H_{30}-\frac{4}{21}m_{0}^{\ast}H_{21}+{}\nn\\
& &{}+\frac{1}{21}m_{1}^{\ast}H_{11}+\frac{1}{21}m_{2}^{\ast}H_{10}.
\eea
%%%%%%%%%%%
One could make the following comments  on the expressions above:
\begin{enumerate}
        \item While the quadratic quantities $M_{ij},Q_{ij}$ are independent quantities,
    $S_{ij},H_{ij}$ are not completely independent of each other.
    The following obvious relations between them hold
    \be
    S_{ij}=-H_{j+1~i-1}.
    \ee
    Thus, although the introduction of four quantities looks like a redundancy this gives
    a  symmetric form in the two sets of moments.
       \item The expressions for the moments coincide with the expressions of \cite{Fodoretal},
    if there is no electromagnetic field.
    \item The different expressions we have in $P_4,Q_4,P_5,Q_5$ moments is due to
    incorrect replacement of $q,\xi$ with $\tilde{q},\tilde{\xi}$ in $s_i$ in Eq.~(\ref{ricci}).
    As was noted in Section \ref{sec:2}, the new expressions for the moments are
    now dimensionally right (see note (v) below).
    \item From the form of the expressions we notice that there is a symmetry between  gravitational
    and  electromagnetic moments. The electromagnetic moments arise from the corresponding
    gravitational moments if we
    interchange $q_i$ with $m_i$,  replace $Q_{ij}$ with $M_{ij}$ and $S_{ij}$ with $H_{ij}$, and
    reverse the signs of all but the single $m_i$, $q_i$ terms.
    \item Since $\xi$ should be dimensionless, $m_i$ and $q_i$ should have dimensions $[M]^{i+1}$,
    and hence all terms in each moment
    have the same (sum of indices) $+$ (number of indices). This simple observation gives us the
    capability of predicting  what kind of terms we expect at each  moment.
    This dimensional analysis makes it clear that the expressions (27,28) of \cite{HoensPerj} contain
    incorrect terms.
    \end{enumerate}

%%%%%%%%%%%%%%%%%%%%%%%%%%%%%%%%%%%%%%%%%%%%%%%%%%%%%%%%%
\section{A spinning mass with charge and magnetic dipole}
\label{sec:4}%%%%%%%%%%%%%%%%%%%%%%%%%%%%%%%%%%%%%%%%%%%%

For completeness, we rewrite the results of the pole-dipole example used by Hoenselaers and Perj\'{e}s,
based on the correct expressions for the moments. Thus,  we also assume that the moments of the source are
%%%%%
\bea
P_{0}=m,  &                &Q_{0}=e,          \nn\\
P_{1}=iam,&                &Q_{1}=i \mu e,    \nn\\
P_{n}=0,  &\textrm{and  ~~}&Q_{n}=0, \textrm{for } n \geq 2 .
\eea
%%%%%%%%%
Then, if we invert the expressions of Eqs.~(\ref{mmelfin},\ref{qqelfin}) we get
%%%%%%%%%%
\bea
m_{0}&=&m\nn\\
m_{1}&=&iam\nn\\
m_{2}&=&0\nn\\
m_{3}&=&0\nn\\
m_{4}&=&\frac{1}{7}am(am^{2}-\mu e^{2})+\frac{3}{70}m\mu e^{2}(a-\mu)\nn\\
m_{5}&=&-i\left[\frac{1}{21}am(a^{2}m^{2}-\mu^{2}e^{2})\right]
\eea
%%%%%%%%%%%
and
%%%%%%%%%%
\bea
q_{0}&=&e\nn\\
q_{1}&=&i\mu e\nn\\
q_{2}&=&0\nn\\
q_{3}&=&0\nn\\
q_{4}&=&\frac{1}{7}\mu e(am^{2}-\mu e^{2})+\frac{3}{70}e a m^{2}(a-\mu)\nn\\
q_{5}&=&-i\left[\frac{1}{21}\mu e(a^{2}m^{2}-\mu^{2}e^{2})\right].
\eea
%%%%%%%%%%%
Although $m_2=q_2=m_3=q_3=0$, all higher order $m_i$ and $q_i$ do not vanish. Especially, if
we set $\mu=a$,
%%%%%%%%%%
\bea
m_{0}&=&m\nn\\
m_{1}&=&iam\nn\\
m_{2}&=&0\nn\\
m_{3}&=&0\nn\\
m_{4}&=&\frac{1}{7}a^{2} m(m^{2}-e^{2})\nn\\
m_{5}&=&-i\left[\frac{1}{21}a^{3} m(m^{2}-e^{2})\right]
\eea
%%%%%%%%%%%
and
%%%%%%%%%%
\bea
q_{0}&=&e\nn\\
q_{1}&=&iae\nn\\
q_{2}&=&0\nn\\
q_{3}&=&0\nn\\
q_{4}&=&\frac{1}{7}a^{2} e(m^{2}-e^{2})\nn\\
q_{5}&=&-i\left[\frac{1}{21}a^{3} e(m^{2}-e^{2})\right],
\eea
%%%%%%%%%%%
which are exactly what the formulae of Hoenselaers and Perj\'es yield.
If on the other hand  we set $e^2=m^2$, then
%%%%%%%%%%
\bea
m_{0}&=&m\nn\\
m_{1}&=&iam\nn\\
m_{2}&=&0\nn\\
m_{3}&=&0\nn\\
m_{4}&=&m^{3}(\frac{1}{7}a+\frac{3}{70}\mu)(a-\mu)\nn\\
m_{5}&=&-i\left[\frac{1}{21}am^{3}(a^{2}-\mu^{2})\right]
\eea
%%%%%%%%%%%
and
%%%%%%%%%%
\bea
q_{0}&=&e\nn\\
q_{1}&=&i\mu e\nn\\
q_{2}&=&0\nn\\
q_{3}&=&0\nn\\
q_{4}&=&e^{3}(\frac{1}{7}\mu+\frac{3}{70}a)(a-\mu)\nn\\
q_{5}&=&-i\left[\frac{1}{21}\mu e^{3}(a^{2}-\mu^{2})\right]
\eea
%%%%%%%%%%%
which do not coincide with the corresponding expressions of Hoenselaers and Perj\'{e}s.
Of course if $e^2=m^2$ and $\mu=a$, then all $q_i$ and $m_i$ except the first pair of them
vanish.

%%%%%%%%%%%%%%%%%%%%%%%%%%%%%%%%
\section{The Kerr-Newman metric}
\label{sec:5}%%%%%%%%%%%%%%%%%%%

In this section we follow the technique presented here to compute
the gravitational and electromagnetic moments of a Kerr-Newman black hole.
Ernst \cite{Ernst2} constructs this metric in prolate spheroidal coordinates $(x,y)$, by setting
\be
\xi_E=\frac{x+i a y}{m} ,\qquad q_E=\frac{e}{m}.
\ee
The $\xi$ we have used in our paper (and in the majority of relevant
papers) is the inverse of Ernst's (compare Eq.~(\ref{Efunction}) of the present paper
with Eq.~(10) of \cite{Ernst1}), and $q=q_E \xi$. Additionally, we will use the opposite sign
for $a$ from the one used by Ernst, since
his convention leads to a spinning black hole with its spin along the $-z$ direction
for positive $a$ (see footnote 24 of Israel \cite{Israel}). Thus,
%%%%%%%%%%%
\bea
\xi=\frac{m}{x-i a y} ,\qquad
q  =\frac{e}{x-i a y}.
\eea
At the axis of the black hole ($\rho=\bar\rho=0$), the prolate spheroidal coordinates are
$y=1$, and $x=z$. Therefore, the Ernst potentials along the axis take the following form
\bea
\xi(\bar{\rho}=0)=\frac{m}{z-i a }=\frac{m\bar{z}}{1-i a \bar{z}} ,\qquad
q(\bar{\rho}=0)  =\frac{e}{m} \xi(\bar{\rho}=0),
\eea
%%%%%%%%%%%
and finally the conformally transformed potentials are
%%%%%%%%%%%%%%
\bea
\tilde{\xi}(\bar{\rho}=0)&=&\frac{m}{1-i a \bar{z}}= m \sum_{l=0}^{\infty} (ia\bar{z})^l \nn\\
\tilde{q}(\bar{\rho}=0)  &=&\frac{e}{1-i a \bar{z}}= e \sum_{l=0}^{\infty} (ia\bar{z})^l .
\eea
Now it is straightforward to read the multipole moments of such a source.
It is easy to see that in this case,
as in the case of Kerr metric, $M_{ij}=Q_{ij}=S_{ij}=H_{ij}=0$, so the $m_{i}$'s and $q_{i}$'s
coincide with the actual moments.
%%%%%%%%%%%%%%%%%%%
\bea
P_{l}&=&m(ia)^{l},\nn\\
Q_{l}&=&e(ia)^{l}.
\eea
%%%%%%%%%%%%%%%%%%%
This set of moments fully defines the gravitational and electromagnetic field of a Kerr-Newman
black hole.

\section*{Acknowledgements}
We would like to thank Prof. Manko for drawing our attention to the odd convention for $a$ used by
Ernst \cite{Ernst1} to describe the Kerr-Newman metric.
This research was supported in part by Grant No 70/4/4056 of the Special Account for Research Grants of the
University of Athens, and in part by the ``PYTHAGORAS'' research funding program.

\end{document}